\documentstyle[aps,pre,twocolumn,epsfig]{revtex}
\newcommand{\bel}[1]{\begin{equation}\label{#1}}
\newcommand{\bra}[1]{\mbox{$\langle \, {#1}\, |$}}
\newcommand{\ket}[1]{\mbox{$| \, {#1}\, \rangle$}}
\newcommand{\exval}[1]{\mbox{$\langle \, {#1}\, \rangle$}}
\newcommand{\abr}[1]{\mbox{$\langle\langle\,#1\,\rangle\rangle$}}
\def\bbbc{{\mathchoice {\setbox0=\hbox{$\displaystyle\rm C$}\hbox{\hbox
to0pt{\kern0.4\wd0\vrule height0.9\ht0\hss}\box0}}
{\setbox0=\hbox{$\textstyle\rm C$}\hbox{\hbox
to0pt{\kern0.4\wd0\vrule height0.9\ht0\hss}\box0}}
{\setbox0=\hbox{$\scriptstyle\rm C$}\hbox{\hbox
to0pt{\kern0.4\wd0\vrule height0.9\ht0\hss}\box0}}
{\setbox0=\hbox{$\scriptscriptstyle\rm C$}\hbox{\hbox
to0pt{\kern0.4\wd0\vrule height0.9\ht0\hss}\box0}}}}
\def\be{\begin{equation}}
\def\ee{\end{equation}}
\def\bea{\begin{eqnarray}}
\def\eea{\end{eqnarray}}
\def\ba{\begin{array}}
\def\ea{\end{array}}
\draft
\begin{document}
\tighten
\onecolumn
\twocolumn[\hsize\textwidth\columnwidth\hsize\csname @twocolumnfalse\endcsname

\title{Exact time-dependent correlation functions for the symmetric
exclusion process with open boundary
}
\date{\today}
\author{
J.E. Santos$^{1}$ and G. M. Sch\"utz$^{2}$}
\address{
$^1$ Physik Department, TU M\"unchen, James-Franck-Strasse, 85747 Garching,
Germany\\
$^2$ Institut f\"ur Festk\"orperforschung, Forschungszentrum J\"ulich,
D-52425 J\"ulich, Germany
}
\maketitle
\begin{abstract}
As a simple model for single-file diffusion of hard core particles we
investigate the one-dimensional symmetric exclusion process. We consider an 
open semi-infinite system where one end is coupled to an external reservoir
of constant density $\rho^\ast$ and which initially is in an non-equilibrium
state with bulk density $\rho_0$. We calculate the exact time-dependent
two-point density correlation function
$C_{k,l}(t)\equiv \exval{n_k(t)n_l(t)} - \exval{n_k(t)}\exval{n_l(t)}$  
and the mean and variance of the integrated average net flux of
particles $N(t)-N(0)$ that have entered (or left) the system up to
time $t$. We find that the boundary region of 
the semi-infinite {\em relaxing} system is in a state similar 
to the bulk state of a finite
{\em stationary} system driven by a boundary gradient. The symmetric
exclusion model provides a rare example where such behavior can be proved
rigorously on the level of  equal-time two-point correlation functions.
Some implications for the relaxational dynamics of entangled polymers and for
single-file diffusion in colloidal systems are discussed.
\end{abstract} 
]
\section{Introduction}
\label{Intro}

The simple symmetric exclusion process (SEP) \cite{Spit70} is a Markov
process describing the diffusive motion of identical hard-core particles on
a lattice. Particles hop randomly to nearest-neighbor sites  with an
exponential waiting-time distribution with mean $\tau_0$, provided the chosen
site is empty. If it is occupied, the attempt to move is rejected.  A
considerable body of exact results, particularly for the one-dimensional case
(see \cite{Ligg85,Schu00} for reviews), has led to a thorough understanding
of many fundamental properties of this model. Applications to specific 
problems include interface growth in the universality class of the
Edwards-Wilkinson equation \cite{Meak86,Plis87}, reptation dynamics of 
entangled polymers \cite{Schu99}, or single-file diffusion in
molecular-sized channels such as biological membrane channels \cite{Chou98}
or zeolite pores \cite{Kukl96}, and in one-dimensional colloidal systems
\cite{Wei00}.

{}From a theoretical and also experimental point of view the main quantities 
of interest are (a) the equilibrium and stationary non-equilibrium properties
of the system with open boundaries where particles can enter and leave
\cite{Spoh83} and (b) the time evolution of the local
particle density and their correlations in a system which relaxes
after it has been prepared in some far-from-equilibrium initial state.
For instance, in the study of kinetic roughening of an initially 
flat interface in a two-phase Ising system one 
obtains from the SEP exact universal scaling functions for
the roughness \cite{Abra94}. 
In the investigation of the reptation dynamics \cite{deGe71,Doi86} of 
an initially stretched DNA chain, flourescence microscopy allows for 
a direct measurement of the relaxation of the tube length \cite{Perk94} which 
in the lattice gas approach \cite{Rubi87}
is proportional to the number of particles in the one-dimensional exclusion
process with open boundary. 
Not only universal power laws, but also non-universal amplitudes are of 
interest, e.g., for estimating the sensitivity of the 
coarse-graining involved in the lattice gas description to the
microscopic properties of
the polymer dynamics in a dense solution \cite{Schu99,Schu01}.

{}For a more detailed understanding of the role of correlations
in non-equilibrium states of the symmetric exclusion process \cite{Diet83}, we
compute here the time-dependent two-point density correlation function in an
one-dimensional semi-infinite system with one open boundary,
connected to a reservoir of constant density $\rho^\ast$.
In the polymer context this equivalent to an
entropic tensile force acting at the end segments of an
entangled polymer chain \cite{Doi86}.

The paper is organized as follows. In order to overcome some of the technical
difficulties connected with the coupling to boundary reservoirs we formulate
the SEP in terms of the dynamical matrix product ansatz
\cite{Stin95a,Stin95b,Sant97} (Sec. II). In Sec. III  we derive a functional
equation for the two-point density correlation function  which is solved by a
Bethe wave function. From this we derive in section IV the scaling form in
terms of error functions. The details of these rather involved calculations
are presented in the appendices. In Sec. V we focus on the fluctuations in the
total number of particles $Q(t)=N(t)-N(0)$ which enter and leave the system
which initially is prepared in an uncorrelated random state with a density
$\rho_0$. We obtain the (expected) universal asymptotic growth law
$\exval{Q^2}-\exval{Q}^2 = A \sqrt{t/\tau_0}$ and the non-universal amplitude
$A(\rho_0,\rho^\ast)$ relevant for reptation dynamics. We conclude with some
brief  remarks on the nature of the evolving non-equilibrium state (Sec. VI).

\section{Dynamical matrix product ansatz}
\label{SEP}

We first consider the symmetric exclusion process on a chain of
$L$ sites. At the boundary sites $k=1,L$ particles are injected (extracted)
with rates $\alpha_{1,L}$ 
($\gamma_{1,L}$). In terms of the transition rates
$w_{\underline{n}\to\underline{n}'}$ from state $\underline{n}$ to
$\underline{n}'$ the stochastic dynamics is described by a master equation 
\bel{2-1}
\frac{d}{dt} P(\underline{n};t) = \sum_{\underline{n}'\in X}
\left[w_{\underline{n}'\to\underline{n}}P(\underline{n}';t) -
w_{\underline{n}\to\underline{n}'}P(\underline{n};t)\right]
\ee
for the probability $P(\underline{n};t)$ of finding, at time $t$, a
configuration $\underline{n}$ of particles on a lattice of $L$ sites. Here
$\underline{n} = \{n_1, n_2, \dots , n_L\}$ where $n_i=0,1$ are the
integer-valued particle occupation numbers at site $i$. 
In what follows it is convenient to set the microscopic time 
unit $\tau_0=1$. In these units 
the single-particle diffusion coefficient 
is given by $D_0=1/2$.

Inserting the transition rates as described above one readily obtains the
equations of motion for $m$-point joint probabilities $\exval{n_{k_1}\dots
n_{k_m}}$. These form a hierarchy of coupled equations where the time
derivatives of the $m$-point joint probabilities are coupled to 
$m-1$-point
joint probabilities. In principle, this allows 
for a recursive solution, which,
however, is  difficult to obtain for $m>1$. 
More importantly, unlike in a periodic
system, the lack of translational invariance 
makes it difficult to
obtain exact results even for $m=2$ 
which we study in this paper. Hence we
reformulate the dynamics in terms of the 
dynamical matrix product ansatz
\cite{Stin95a,Stin95b,Sant97}. This ansatz 
leads to a decoupling of the joint probabilities 
and splits the many-body dynamical problem
into a (trivial) dynamical single-particle 
part and a (non-trivial) many-body
part which is time-independent 
and which can be solved using the Bethe ansatz.

The dynamical matrix product ansatz is reviewed in detail in \cite{Schu00}.
Here we discuss only the main features relevant for the SEP.  One represents
an occupied (vacant) site by a time-dependent matrix $D$ ($E$) in a string 
$DDDEDEE\dots$ of $L$ such matrices. The configurational probabilities 
$P(\underline{n};t)$ are obtained by sandwiching the product of these $L$
matrices $D$ or $E$ between suitably chosen vectors $\langle\bra{W}$ and
$\ket{V}\rangle$ and normalizing by $Z_L=\langle\bra{W}C^L\ket{V}\rangle$
where $C=D+E$. [Notice that expanding  the $L$-th power of $C$ automatically
gives the sum of all unnormalized configurational probabilities and hence
yields the correct normalization factor.] 
The time-dependent matrices satisfy
algebraic relations which are determined by 
requiring the matrix product state
to satisfy the master equation (\ref{2-1}). Expectation values of local
observables are obtained by sandwiching suitable products of the matrices
$D,E$ with $C$. Defining formally $D_k=C^{k-1}DC^{-k}$ one obtains for the
local particle density $\rho_k(t)=\exval{n_k(t)}$ at site $k$ 
\bea \rho_k(t) &
\equiv & \langle\bra{W}C^{k-1}DC^{L-k}\ket{V}\rangle/Z_L\\  
\label{2-10}
 & = &  \langle\bra{W}D_kC^{L}\ket{V}\rangle /Z_L
\eea
and for the joint probability $G_{k,l}(t)=\exval{n_k(t)n_l(t)}$ of finding
particles at sites $k,l$ 
\bel{2-11}
G_{k,l}(t) =  \langle\bra{W}D_kD_lC^{L}\ket{V}\rangle /Z_L.
\ee
{}From these quantities one obtains the two-point density
correlation function
\bel{2-6}
C_{k,l}(t)= \exval{n_k(t)n_l(t)} - \exval{n_k(t)} \exval{n_l(t)}.
\ee
Higher-order joint probabilities are obtained analogously.
The initial probability distribution is encoded in the matrices
$D(0)$.

We do not review here how the matrix relations and the corresponding
relations for the vectors $\langle\bra{W}$, 
$\ket{V}\rangle$ are obtained from
the master equation, but refer the reader 
to \cite{Schu00}. The matrix $C$ as 
well as the vectors $\langle\bra{W}$ and 
$\ket{V}\rangle$ may be chosen to be
time-independent \cite{Sant97}. 
The dynamical problem is
then solved by introducing 
Fourier transforms 
\bel{2-12}
{\cal D}_p(t) = \sum_k e^{ipk} D_k(t).
\ee
They have the simple time-dependence
\be\label{2-13}
{\cal D}_p(t) = e^{-\epsilon_p t} {\cal D}_p(0)
\end{equation}
in terms of the initial matrix ${\cal D}_p(0)$ and 
the inverse relaxation 
times $\epsilon_p=1-\cos{p}$. These 
diffusive single-particle relaxation modes
reflect the random walk nature of the dynamics. For calculating the local
expectation values (\ref{2-10}), (\ref{2-11}) it is useful to separate the
static ($p=0$) and dynamical ($p\neq 0$) parts in the Fourier expansion of
$D_k$ and to write the inverse Fourier transform in the form 
\begin{equation}\label{2-15}
D_k(t) = (1-k){\cal D}_{0} + {\cal I} + 
 \int' \frac{dp}{2\pi}\, {\cal D}_{p}(0)\,
e^{-ipk-\epsilon_p t}\,.
\end{equation}
In the semi-infinite system 
which we will be considering,
the primed integral is to be understood as a contour integral in the
variable $z=e^{-ip}$ where the contour is chosen in a way such
that the values of the joint probabilities obey the correct
initial conditions at $t=0$.
The matrices ${\cal D}_{0}$ and ${\cal I}$ are 
time-independent and yield all
stationary expectation values. The time-dependent 
integral contains the relaxational part of $D_k$.

The master equation not only determines the 
time evolution of ${\cal D}_p(t)$, which is given
by (\ref{2-13}), but also
requires the Fourier components of 
$D_k(t)$, as given by (\ref{2-15}), to satisfy various
relations among themselves 
and with the vectors $\langle\bra{W}$, $\ket{V}\rangle$.
The dynamical part where both $p_1$ and $p_2$ are non-zero satisfies
\bel{2-17}
{\cal D}_{p_1}{\cal D}_{p_2} = S(p_1,p_2) {\cal D}_{p_2}{\cal D}_{p_1} 
\ee
with
\begin{equation}\label{2-18}
S(p_1,p_2) = - \frac{1+e^{i p_1 + ip_2}-2e^{i p_2}}
{1+e^{i p_1 + ip_2}-2e^{i p_1}}.
\end{equation}
The relations involving static components read
\bea\label{2-19}
\left[\,{\cal D}_p\,,\,{\cal D}_0\,\right] & = & 0\\
\label{2-20}
\left[\,{\cal D}_p\,,\,{\cal I}\,\right] & = & 2 {\cal D}_0{\cal D}_p\\
\label{2-21}
\left[{\cal D}_0\,,\, {\cal I}\,\right] & = & {\cal D}_0^2.
\eea
These relations have their origin in the bulk exclusion interaction 
between particles.

The boundary conditions determine the action of ${\cal D}_p$ on
the vectors $\langle\bra{W}$ and $\ket{V}\rangle$. One finds
\begin{eqnarray}
\label{2-23}
0 & = & \langle \bra{W} \left\{ {\cal D}_{0} +
        2(\alpha_1 + \gamma_1){\cal I} - 2\alpha_1  \right\} \\
\label{2-24}
0 & = & \langle \bra{W} \left\{ {\cal D}_{p} +
        \mbox{e}^{2ip}B_1(p){\cal D}_{-p}) \right\}
        \hspace{6mm} (p \neq 0) \\
\label{2-25}
0 & = & \left\{ (2\alpha_L + 2\gamma_L - 1){\cal D}_{0} +
        2(\alpha_L + \gamma_L ){\cal I} - 2\alpha_L \right\} \ket{V}\rangle\\
\label{2-26}
0 & = & \left\{ B_L(p){\cal D}_{p} +
        {\cal D}_{-p} \right\} \ket{V} \rangle
        \hspace{6mm} (p \neq 0) .
\end{eqnarray}
with
\be\label{2-22}
B_i(p) = \frac{2\alpha_i+2\gamma_i-1+\mbox{e}^{-ip}}
{2\alpha_i+2\gamma_i-1+\mbox{e}^{ip}}.
\ee
Notice that in the relations (\ref{2-17}) - (\ref{2-26}) the time-dependence
drops out. The set of relations (\ref{2-13}),  (\ref{2-17}) -  (\ref{2-22})
provides an alternative mathematical formulation of the symmetric exclusion
process with open boundaries. 

\section{Correlation functions}
\label{corrfunc}

{}For the equilibrium choice of boundary parameters $\alpha\equiv\alpha_1 =
\alpha_L = \rho^\ast/2$ and $\gamma\equiv\gamma_1=\gamma_L=(1-\rho^\ast)/2$
the boundary relations (\ref{2-23}) - (\ref{2-26}) and the functions $B_i(p)$
simplify considerably. With these rates, modelling the connection to 
particle reservoirs of density $\rho^\ast$, the (unique) invariant measure of
the process is a product measure with density $\rho^\ast$, i.e., there are no
density correlations between different sites. Choosing as initial state an
uncorrelated state with density $\rho_0\neq\rho^\ast$ leads to a non-trivial
time evolution as the system starts to fill up ($\rho_0<\rho^\ast$) 
or deplete (respectively $\rho_0>\rho^\ast$). Correlations 
are build up in the transient regime before the equilibrium state is
attained. In a semi-infinite  system this will
take an infinite amount of time, and the ``transient'' regime is the only
relevant one. For finite systems with $L$ sites \cite{Spoh83} the system is
transient for times $t < \tau^\ast \propto L^2$. This is the regime on which
we focus our attention. Therefore we ignore the right boundary site by
considering the thermodynamic limit $L\to\infty$. The only remaining
length scale (besides the unit lattice constant) is then the dimensionless
diffusion length
\be\label{3-1b}
\tilde{L} \equiv \tilde{L}(t) = \sqrt{4t/\pi\tau_0}\,.
\ee 
This is a dynamical length scale playing the 
role of a correlation length (see below).

Anticipating the importance of macroscopic {\em static} initial and 
equilibrium properties for the non-equilibrium relaxation 
process we introduce
the basic quantities characterizing both the non-equilibrium initial state
with density $\rho_0$ and the asymptotic equilibrium state  with density
$\rho^\ast$.  These are the density gradient 
\bel{3-2}
\Delta\rho=\rho^\ast - \rho_0
\ee
between bulk and boundary and the compressibility 
\be
\kappa = \lim_{L\to\infty}(\exval{N^2}-\exval{N}^2)/L
\ee
which is readily obtained from the static two-point 
correlation function
\bel{2-9}
C_{k,l} = \rho(1-\rho)\delta_{k,l}
\ee
of the uncorrelated initial and final distributions respectively.
Here $\delta_{k,l}$ is the Kronecker delta-function. Hence
\bel{2-9b}
\kappa = \rho(1-\rho)
\ee
where $\rho=\rho_0$ or $\rho^\ast$ respectively.

\subsection{Density profile and current}

The evolving density profile was computed 
exactly in Ref. \cite{Stin95a}. If $\alpha_1=\alpha_L=\rho^\ast/2$, 
$\gamma_1=\gamma_L=(1-\rho^\ast)/2$, one has
\begin{eqnarray}
\rho_k(t)&=&
\rho^\ast +
\int^{'}\,\frac{dp}{2\pi}\,\abr{
{\cal D}_{p}}\,e^{-ipk-\epsilon_p t}\,,
\label{3-01}
\end{eqnarray}
where we use the abbreviation
\begin{equation}
\label{3-02}
\abr{{\cal D}_{p_1}\ldots{\cal D}_{p_N}}\;\equiv\;
\frac{\langle\bra{W}\,{\cal D}_{p_1}\ldots{\cal D}_{p_N}\,C^{L}
\,\ket{V}\rangle}{Z_L}\,,
\end{equation}
in order to make the formulas more compact.
Using equations (\ref{2-24}) and (\ref{2-26})
with $B_1(p)=B_L(p)=e^{-2ip}$, one
can show that $\abr{{\cal D}_{p}}$ obeys the functional
equation  $\abr{{\cal D}_{p}}=-\abr{{\cal D}_{-p}}$,
with the solution
\begin{equation}
\label{3-03}
\abr{{\cal D}_{p}}\;=\;\sum_{k_0=1}^{L}\,a_{k_0}\,
(\,e^{ipk_0}-e^{-ipk_0}\,)\,,
\end{equation}
where the constants $a_{k_0}$ are determined by the initial
conditions. Furthermore, equations (\ref{2-24}) and (\ref{2-26})
impose constraints on the set of allowed momenta, given by 
$e^{2ip(L+1)}=1$. In the thermodynamic limit $L\rightarrow
\infty$, the momenta $p$ form a continuous set and this
condition can be relaxed. For a system which is initially
in an uncorrelated state with density $\rho_0$, one
has $a_{k_0}=-\Delta\rho$ and one obtains
in the thermodynamic limit, after substitution of
(\ref{3-03}) in (\ref{3-01})     
\begin{equation}
\label{3-1}
\rho_k(t) = \rho_0 + (\rho^\ast - \rho_0)\,g_k(t)\,,
\end{equation}
where $g_k(t)$ is the lattice analog of 
the complementary error function
\bel{3-1c}
g_k(t) = 
\mbox{e}^{-t}\left[I_k(t)+
2\sum_{p=k+1}^\infty I_p(t)\right]
\ee
and where $I_k(t)$ are 
the modified Bessel functions (\ref{A-0}). 
In terms of the scaling variable 
\bel{3-11}
\tilde{x} = k/\tilde{L}
\ee
the long-time behavior of the density
profile is given by the error function, as is well-known for diffusive
transport. In the vicinity of the boundary, i.e., at distances $\tilde{x} \ll
1$ small compared to the diffusion length, the density profile is linear.

Associated with the spatial variation of the density
there is a diffusive relaxational current $j_k = D_0(\rho_k(t) -
\rho_{k+1}(t))$ which is space-independent close to the boundary
to lowest leading order in time. It is convenient to define
the current 
\be
\hat{\jmath} = j/D_0
\ee 
in units of the single-particle diffusion coefficient $D_0=1/(2\tau_0)$. From
the expansion (\ref{A-1}) of the Bessel function one finds  
\bea 
\hat{\jmath} & = & 2 (\rho^\ast - \rho_0) /\sqrt{2\pi t} \nonumber\\ 
\label{3-1a}
 & = & 2\sqrt{2} \Delta\rho/(\pi\tilde{L}).
\eea

\subsection{Two-point correlation function: Exact expression}

The density profile $\rho_k(t)$ and hence the time-dependent {\em equilibrium}
two-point correlation functions $C^\ast_{k,l}(t) = \exval{n_k(t)n_l(0)} -
(\rho^\ast)^2$ can be obtained in a straightforward manner from the solution
of a lattice diffusion equation. The solution of the equations of motion for
equal-time joint probabilities, however, and hence the calculation of the
time-dependent two-point correlation function (\ref{2-6}) is much more
involved. A convenient way to circumvent an explicit integration of the
coupled equations is to make use of the algebraic representation of
expectation values within the dynamical matrix product ansatz. Substituting
equation (\ref{2-15}) in (\ref{2-11}) and using the commutation
relations (\ref{2-19}-\ref{2-21}), one obtains, when  
$\alpha_1=\alpha_L=\rho^\ast/2$, $\gamma_1=\gamma_L=(1-\rho^\ast)/2$,
\begin{eqnarray}
\exval{n_k(t)\,n_l(t)}&=&\rho^{\ast\,2}+\rho^{\ast}\,
\int^{'}\,\frac{dp}{2\pi}\,\abr{{\cal D}_{p}}\,e^{-ipk-\epsilon_p
t}\nonumber\\
& &\mbox{}+
\rho^{\ast}\,\int^{'}\,\frac{dp}{2\pi}\,\abr{{\cal D}_{p}}\,e^{-ipl-\epsilon_p
t}\nonumber\\
& &\mbox{}+\int^{'}\,\int^{'}\,\frac{dp_1}{2\pi}\,\frac{dp_2}{2\pi}\,
\abr{{\cal D}_{p_1}\,{\cal D}_{p_2}}\nonumber\\
& &\mbox{}\times
e^{-i(p_1k+p_2l)-(\epsilon_{p_1}+
\epsilon_{p_2})t}\,,
\label{3-3}
\end{eqnarray} 
where $\abr{{\cal D}_{p}}$ is given by (\ref{3-03}).

Using equations (\ref{2-17}) and (\ref{2-24}), we can show
that $\abr{{\cal D}_{p_1}\,{\cal D}_{p_2}}$ obeys
the following functional equation
\begin{equation}
\label{3-3a}
\abr{{\cal D}_{p_1}\,{\cal D}_{p_2}}=-
S(-p_1,p_2)\,\abr{{\cal D}_{p_2}\,{\cal D}_{-p_1}}\,,
\end{equation}
together with similar equations involving 
all the possible arrangements
of $p_1$ with $p_2$ or $-p_2$ and $-p_1$ 
with $p_2$ or $-p_2$. Furthermore, in a finite
system, these equations and equation (\ref{2-26})
determine the set of allowed momenta. These equations
are the equations obeyed by the Bethe wave function of
a quantum spin $1/2$ system with boundary fields 
\cite{Alca87} and they have the solution
\begin{equation}
\label{3-3b}
\abr{{\cal D}_{p_1}\,{\cal D}_{p_2}}=\sum_{k_0<l_0}^{L}a_{k_0,l_0}
\Psi_{p_1,p_2}(k_0,l_0)
\end{equation}
where the constants $a_{k_0,l_0}$ are determined by the initial
conditions and $\Psi_{p_1,p_2}(k_0,l_0)$ is the Bethe wave function,
which is given by
\begin{eqnarray}
\label{3-4}
\Psi_{p_1,p_2}(k_0,l_0)&=&\,
\,e^{ip_1k_0+ip_2l_0}
\,+\,S(p_1,p_2)\,
e^{ip_2k_0+ip_1l_0}\\
& & \mbox{}-e^{-ip_1k_0+ip_2l_0}
-S(-p_1,p_2)\,e^{ip_2k_0-ip_1l_0}\nonumber\\
& &\mbox{}-S(p_1,p_2)\,e^{-ip_2k_0+ip_1l_0}\nonumber\\
& &\mbox{}+S(-p_1,p_2)\,e^{-ip_2k_0-ip_1l_0}\nonumber\\
& &\mbox{}-
S(-p_1,p_2)\,S(p_1,p_2)\,e^{ip_1k_0-ip_2l_0}\nonumber\\
& &\mbox{}+
S(-p_1,p_2)\,S(p_1,p_2)\,e^{-ip_1k_0-ip_2l_0}\,.\nonumber
\end{eqnarray}
Using equation (\ref{3-01}), one can write the two-point
correlation function as
\begin{eqnarray}
\label{3-5}
C_{k,l}(t)&=& \int^{'}\,\int^{'}\,\frac{dp_1}{2\pi}\,\frac{dp_2}{2\pi}\,
(\, \abr{{\cal D}_{p_1}\,{\cal D}_{p_2}}\\
& &\mbox{}-\abr{{\cal D}_{p_1}}
\abr{{\cal D}_{p_2}}\,)\,
e^{-i(p_1k+p_2l)-(\epsilon_{p_1}+
\epsilon_{p_2})t}\nonumber
\end{eqnarray}
where $\abr{{\cal D}_{p_1}\,{\cal D}_{p_2}}$ is given
by (\ref{3-3b}) and $\abr{{\cal D}_{p_1}}$ by (\ref{3-03}).

If we now choose the contour of integration such that 
at $t=0$ the integral over $\Psi_{p_1,p_2}(k_0,l_0)$ in (\ref{3-5})
is equal to $\delta_{k,k_0}\,\delta_{l,l_0}$,
then the condition that the initial state is 
uncorrelated $C_{k,l}(0)=0$, gives 
$a_{k_0,l_0}=a_{k_0}\,a_{l_0}=(\Delta\rho)^2$. 
Substituting this result in (\ref{3-3b}) and using
this equation, together with (\ref{3-03}) in (\ref{3-5}),
we obtain for the semi-infinite system, when $L\rightarrow\infty$,
\begin{eqnarray}
\label{3-6}
C_{k,l}(t)&=&(\Delta\rho)^2\left\{
\,e^{-2t}\,(\,I_{l+k}(2t)-I_{l-k}(2t)\,)\right.\\
& &\mbox{}+\sum_{k_0<l_0}^{\infty}\,
\int^{'}\,\int^{'}\,\frac{dp_1}{2\pi}\,\frac{dp_2}{2\pi}\,
e^{-i(p_1k+p_2l)-(\epsilon_{p_1}+\epsilon_{p_2})t}\nonumber\\
& &\mbox{}\times
\left[\,(S(p_1,p_2)-1)\,e^{ip_2k_0+ip_1l_0}\right.\nonumber\\
& &\mbox{}-(S(-p_1,p_2)-1)\,e^{ip_2k_0-ip_1l_0}\nonumber\\
& &\mbox{}-(S(p_1,p_2)-1)\,e^{-ip_2k_0+ip_1l_0}\nonumber\\
& &\mbox{}+(S(-p_1,p_2)-1)\,e^{-ip_2k_0-ip_1l_0}\nonumber\\
& &\mbox{}-(S(-p_1,p_2)\,S(p_1,p_2)-1)\,e^{ip_1k_0-ip_2l_0}\nonumber\\
& &\mbox{}\left.\left.
+(S(-p_1,p_2)\,S(p_1,p_2)-1)\,e^{-ip_1k_0-ip_2l_0}\,
\right]\right\}\,,\nonumber
\end{eqnarray}
where the first term comes from the summation terms of
$\abr{{\cal D}_{p_1}}\abr{{\cal D}_{p_2}}$ with $k_0=l_0$.
We remark that for non-interacting particles (no hard-core repulsion and hence
no exclusion, but otherwise identical hopping dynamics) the correlation
function has only this term, but with a different amplitude $a_0 =
\exval{n^2}_0-\rho_0^2-\rho_0$ determined by the initial
distribution only. The exclusion interaction gives rise to the double
sum over $k_0,l_0$ with the terms of the form $S-1$, $SS'-1$.

To extract more detailed information from the exact expression (\ref{3-6}),
one needs to perform the double sum over $k_0, l_0$ and to
determine the contour of integration such that (\ref{3-6})
obeys the initial condition. This is shown in Appendix~\ref{Bethe}.
As a result, we obtain the more compact expression 
\bea
C_{k,l}(t) & = & -(\Delta\rho)^2 \left[ F_{k+l-1,l-k}(t) +
F_{k+l,l-k+1}(t) \right.\nonumber \\
\label{3-7}
 &  & \left. - F_{l-k-1,k+l}(t) - F_{l-k,k+l+1}(t)\right]
\eea
with
\bea
F_{m,n}(x) & = &  \frac{4\mbox{e}^{-2x}}{\pi} 
\int_0^{\pi/2}d\theta \cos{\theta}\int_0^{x}dv\,\mbox{e}^{2v\cos^2{\theta}} 
\times \nonumber \\
\label{3-8}
 & & I_m(2(x-v)\cos{\theta})
\cos{[v\sin{(2\theta)}+n\theta]}\,,
\eea
which will be used in the next section to obtain the scaling
behaviour of the two-point correlation function
and in section \ref{secV} to determine the particle
number fluctuations.

\section{Scaling behaviour of the two-point correlation function}

The expression (\ref{3-7}) is our starting point for analyzing the
particle number fluctuations (see next section). For investigating
local correlations it is more convenient to write (\ref{3-7}) in another
way
\bea
C_{k,l}(t) & = &  -(\Delta\rho)^2 \left[ \hat{F}_{k,l-1}(t) +
\hat{F}_{k,l}(t) \right.\nonumber \\
\label{3-9}
 &  & \left.- \hat{F}_{-k,l-1}(t) - 
\hat{F}_{-k,l}(t)\right]
\eea
with
\bea
\hat{F}_{m,n}(x) & = & 2t \int_0^{1}dv \, \mbox{e}^{-v^2t}
\mbox{e}^{-2t(1-v)} \times \nonumber \\
\label{3-10}
 & & (1-v)^{n-m+1}I_m(t(1-v))I_n(t(1-v)). 
\eea

One can show that (\ref{3-9}) holds if we use the integral representation
(\ref{A-0}) of the modified Bessel functions appearing in the definition
of $F_{m,n}(x)$, as given by (\ref{3-8}). We can then perform the 
integral over $v$. Representing the integral over $\theta$ as a complex
integral over an appropriate contour we then obtain, 
after performing some expansions and using the identities 
(\ref{A-13}) and (\ref{A-7}), $F_{m,n}(x)=
\hat{F}_{{(m-n+1)}/{2},{(m+n-1)}/{2}}(x)$
which, when substituted in (\ref{3-7}), yields (\ref{3-9}).

{}For $t\gg 1$ the main contribution to the integral comes from small
values of $v$. In order to obtain the scaling behavior we define in analogy
to (\ref{3-11}) a second scaling variable $\tilde{y} = l/\tilde{L}$
and substitute the integration variable $u=v\tilde{L}$. To leading order
in time one then has $(1-v)^{k-j} = \exp{[-u(\tilde{x}-\tilde{y})]}$. With
(\ref{A-1}) we obtain the scaling form of the correlation function
\bel{3-12}
C_{\tilde{x},\tilde{y}}(t) = - (\Delta\rho)^2 \sqrt{\frac{\pi}
{t}}  [R(\tilde{x},\tilde{y}) - R(-\tilde{x},\tilde{y})]
\ee
where
\bel{3-13}
R(\tilde{x},\tilde{y}) = 
\mbox{e}^{-(\tilde{x}+\tilde{y})^2/\pi}
 \mbox{erfc}{((\tilde{y}-\tilde{x})/\sqrt{\pi})}.
\ee
and $\mbox{erfc}(.)$ is the complementary error function (Fig.~1). As time
increases, the range of correlations increases proportionally to $\sqrt{t}$,
but the amplitude decreases in the same manner. The negative sign
of the correlation function signals anticorrelations typical for
the exclusion effect \cite{Ligg85}. In a finite system with a fixed density
gradient between the two boundaries imposed by the coupling to two different
particle reservoirs these anticorrelations persist \cite{Spoh83}. They
extend over the whole lattice and have an amplitude inversely proportional
to the system size.

\setlength{\unitlength}{1.0cm}
\begin{figure}
\begin{center}
\epsfig{width=9\unitlength,
       angle =0,
      file=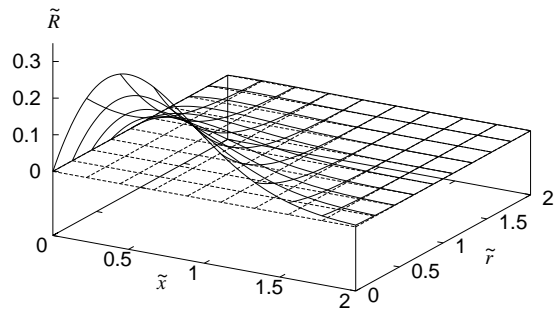}\vspace{3mm}
\end{center}
\caption{Scaling part $\tilde{R} = R(\tilde{x},\tilde{y}) -
R(-\tilde{x},\tilde{y})$ of the two-point correlation function as a function 
of the scaled bulk coordinate $\tilde{x}$ and the scaled lattice distance 
$\tilde{r}=\tilde{y}-\tilde{x}$ (full lines). For visualization purposes
also the plane $\tilde{R}=0$ (equilibrium case) is shown (broken lines).}
\label{F1}
\end{figure}

In the vicinity of the boundary ($0\leq \tilde{x}<\tilde{y}\ll 1$) one has 
$R(\tilde{x},\tilde{y}) - R(-\tilde{x},\tilde{y}) =
4\tilde{x}(1-\tilde{y})/\pi$. In terms of the current (\ref{3-1a})
and the diffusion length $\tilde{L}(t)$ (\ref{3-1b}) the boundary correlation
function is thus given by
\bel{3-14}
C = - \tilde{L} \hat{\jmath}^2 \tilde{x}(1-\tilde{y}).
\ee
Corrections, which can easily be
obtained from the exact scaling function (\ref{3-12}), are of third
order in the scaling variables.

In order to disentangle the effects caused by the initial distribution
and the exclusion interaction, a comparison with non-interacting 
particles is again instructive. The scaling form of the correlation
function is similar
in structure,  but one has $R(\tilde{x},\tilde{y}) = 
\mbox{e}^{-(\tilde{x}-\tilde{y})^2/\pi}$ and a different amplitude
$a_0/\sqrt{4\pi t}$ (see remark after Eq. (\ref{3-6})). For the same
uncorrelated initial distribution as considered for the
exclusion process one has $a_0 = - \rho_0^2 <0$. Remarkably, 
anticorrelations develop even though there is no exclusion. However,
the amplitude is different and,
unlike in the exclusion process, these anticorrelations vanish in a
finite system driven by a boundary gradient as time tends to infinity.

\section{Particle number fluctuations}
\label{secV}

We define as $Q(t) = N(t) - N(0)$ the net number of particles that have
entered or left the system until time $t$.
The mean $\exval{Q(t)}$ is evaluated 
using (\ref{A-2}), (\ref{A-3}), (\ref{A-5}) and one gets
\bea
\exval{Q(t)} & = & \Delta\rho\sum_{k=1}^\infty g_k(t)\nonumber \\
 & = & \Delta\rho\mbox{e}^{-t}\sum_{k=1}^\infty (2k-1) I_k(t)\nonumber  \\
\label{4-1}
 & = & \Delta\rho \left[t \mbox{e}^{-t}(I_0(t)+I_1(t)) - 
\frac{1-\mbox{e}^{-t}I_0(t)}{2} \right].
\eea
The mean grows asymptotically with the power law
\bel{4-2}
\exval{Q(t)} \sim \Delta\rho\sqrt{\frac{2t}{\pi}}
\ee
characteristic for diffusive processes.

The variance
\bel{4-3}
\sigma^2(t)=\exval{Q^2(t)}-\exval{Q(t)}^2
\ee 
may be split into three different parts
\bea
\sigma^2(t)& = & \exval{N^2(t)}-\exval{N(t)}^2 + \exval{N^2(0)}-\exval{N(0)}^2
\nonumber \\
\label{4-4}
 & &  - 
2(\exval{N(t)N(0)}-\exval{N(t)}\exval{N(0)}).
\eea
Since the initial state is a product state one has 
$\exval{N(t)N(0)}-\exval{N(t)}\exval{N(0)}=\kappa_0 \frac{d}{d\rho_0}
\exval{N(t)}$ where $\kappa_0$ (see (\ref{2-9b}))
is the compressibility of the system in the initial state. 

\begin{figure}
\begin{center}
\epsfig{width=9\unitlength,
       angle =0,
      file=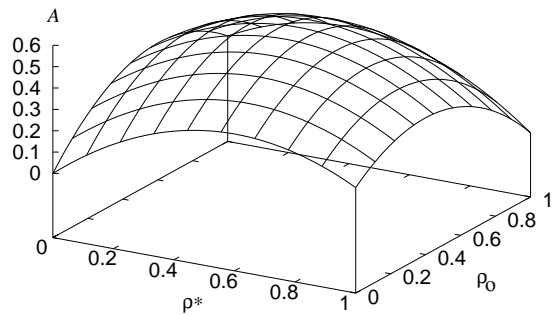}\vspace{3mm}
\end{center}
\caption{Amplitude $A$ of the particle number fluctuations as a function
of the boundary density $\rho^\ast$ and the initial bulk density $\rho_0$.
 }
\label{F2}
\end{figure}

On the other hand, because of the exclusion principle, one can write
\be
\exval{N^2(t)}-\exval{N(t)}^2 = -(\Delta\rho)^2 K(t)
 + \sum_{k=1}^\infty (\rho_k(t)-\rho^2_k(t))
\ee
with the double sum
\be
\label{4-4a}
-(\Delta\rho)^2 K(t) \equiv 2 \sum_{k=1}^\infty\sum_{l=k+1}^\infty 
C_{k,l}(t).
\ee
Using 
\be
\frac{d}{d\rho_0} \exval{n_k(t)} = 1-g_k(t)
\ee
we rewrite (\ref{4-4}) in the form
\bea
\sigma^2(t) & = & \sum_{k=1}^\infty
[(\kappa_0+\kappa^\ast+(\Delta\rho)^2)\,g_k(t)-(\Delta\rho)^2\,g^2_k(t)]
\nonumber \\
\label{4-5}
& & -(\Delta\rho)^2 K(t)
\eea
convenient for studying its asymptotic behavior. The sum over $g_k$
has been calculated above (\ref{4-1}). The
evaluation of the other sums is rather technical, the details are presented
in Appendix~\ref{Laplace}. One finds the following asymptotics \bea
\label{4-5a}
K(t) & \sim & (3-2\sqrt{2})\sqrt{\frac{4t}{\pi}} \\
\label{4-5b}
\sum_{k=1}^\infty g^2_k(t) & \sim & (\sqrt{2}-1)\sqrt{\frac{4t}{\pi}}
\eea
and therefore
\bel{4-6}
\sigma^2(t) = A(\rho_0,\rho^\ast) \sqrt{t}
\ee
with
\bel{4-7}
A(\rho_0,\rho^\ast) = \sqrt{\frac{2}{\pi}}
\left(\kappa_0+\kappa^\ast+(3-2\sqrt{2})
(\Delta\rho)^2\right). 
\ee
The amplitude $A$ is symmetric under interchange of the
macroscopic quantities $\rho_0,\rho^\ast$ and convex in the physical domain
$0\leq \rho_0,\rho^\ast,\leq 1$ with a local maximum at
$\rho_0=\rho^\ast=1/2$ (Fig.~\ref{F2}). {}For an
initially completely filled lattice ($\rho_0=1$) we recover the result
presented previously \cite{Schu01}. For the trivial case of non-interacting
particles one has $(\sqrt{2}-1)\rho_0^2$ instead of the gradient term in
(\ref{4-7}).

\setlength{\unitlength}{1.0cm}

Within the Rubinstein model for reptation \cite{Rubi87} and its extension
\cite{Schu99} the particle number of the symmetric exclusion process is
proportional to the tube length of a polymer and hence proportional to the
experimentally accessible visual length of a flourescence marked entangled
macromolecule such as DNA. An non-equilibrium initial density $\rho_0 >
\rho^\ast$ corresponds to a stretched conformation which may be approximated
by dragging a molecule through a dense solution with optical tweezers
\cite{Perk94}. It has been shown that the relaxing tube length calculated 
from the symmetric exclusion process is in good agreement with experimental
data in the universal initial-time regime \cite{Schu99}.
The particle number fluctuations (the lattice sum over the two-point
density correlation function) that we obtain yield the evolution of the tube
length fluctuations of the polymer chain. For an initially fully stretched
polymer ($\rho_0=1$) the result (\ref{4-6}), (\ref{4-7}) has been discussed
in a recent publication \cite{Schu01}. We see here that a partially
stretched chain displays qualitatively similar relaxational behavior.
In contrast to the Langevin approach used in Rouse-based standard reptation
theory \cite{Doi86} where $A$ depends solely on the equilibrium
tube length fluctuations, the exclusion model predicts a dependence also on 
the initial state (via the compressibility $\kappa_0$) and the amount
of stretching $\Delta\rho$. These features arise from the static interactions
between particles (hard-core, in our case) which are neglected in the purely
entropic Rouse model.

\section{Conclusions}

The main results of our study are the exact expression (\ref{3-9}) and
scaling form (\ref{3-12}) respectively of the time-dependent two-point
density correlation function and the asymptotic variance (\ref{4-6})
in the number of particles that have left or entered the system up to
time $t$. It turns out that the non-equilibrium behavior of the
model is largely determined by the dynamical diffusion length $\tilde{L}$ and
by three static macroscopic  quantities, viz. the compressibilities
$\kappa_0,\kappa^\ast$ of the initial and equilibrium states resp., and the
density gradient $\Delta\rho$ between the (equilibrium) boundary and 
the (initial) bulk density. With regard to polymer reptation this result
(\ref{4-7}) supports our previous conclusion \cite{Schu01} that standard
reptation theory is an oversimplified model of the relaxation
process of stretched, entangled polymers. 

On a local level 
we find an inverse relationship between the range $\xi\propto\tilde{L}$ of
(anti-)correlations and their strength $\propto 1/\tilde{L}$. It is 
interesting to quantitatively compare the expression (\ref{3-14}) for the
correlation function in the boundary region of the system with the
{\em stationary} non-equilibrium correlation function $C^\ast$ of a finite
system of $L$ sites with two different reservoir densities $\rho^- =
\rho^\ast$, $\rho^+ = \rho_0$ as considered in the Section II. The external
density gradient imposes a stationary current  which is given by 
$\hat{\jmath}^\ast = \Delta\rho/L$. In terms of the scaling variables $x=k/L$,
$y=l/L$ one finds $C^\ast = - L (\hat{\jmath}^\ast)^2 x(1-y)$ 
\cite{Spoh83,Sant97} which is of the same form as (\ref{3-14}). This result
suggests that the state of the open system in the boundary region (i.e., at
distances small compared to the diffusion length $\tilde{L}$) is similar to
the non-equilibrium steady state of a finite system of size $L=\tilde{L}$.
Hence we can identify three distinct length scales where the system displays
different behavior. On the scale of the lattice constant $a=1$ the system (in
the boundary region) is in local equilibrium  as is the bulk state of the
finite stationary system \cite{Spoh83}. On intermediate scales $a\ll r \ll
\tilde{L}$ the system is locally (i.e., in the boundary region) stationary, 
but not in equilibrium. On large scales $r\gg\tilde{L}$
the system is neither in equilibrium nor stationary, but displays relaxational
behavior and dynamical scaling.

It is no surprise that the qualitative features of the relaxation process in
simple symmetric exclusion can be described in terms of dynamical scaling with
the diffusion length $\tilde{L}(t)$ and the universal power law  $\sqrt{t}$
characteristic for diffusive dynamics. Yet it is gratifying to have a simple,
but non-trivial model, where not only scaling theories can be verified
explicitly, but also scaling functions can be calculated. An interesting
open problem remains the question whether these results can be obtained
from more widely applicable coarse-grained hydrodynamic approaches 
\cite{Spoh83} for stochastic interacting particle systems. A direct 
experimental study of the questions addressed here appears to be feasible
by studying colloidal particles in a set-up similar to that used in
\cite{Wei00}. In such an experiment the influence of direct particle
interactions in addition to pure hard-core repulsion can be studied.

\section*{Acknowledgments}

J.E.S. would like to thank A. Parmeggiani for useful discussions and
the Institut f\"ur Festk\"orperforschung at FZ J\"ulich for kind hospitality.
J.E.S. acknowledges financial support during the 
different stages of this work from the EU in the
framework of the Contract ERB-FMBI-CT 97-2816 and the DFG in the
framework of the Sonderforschungsbereich SFB 413/TP C6.

\appendix
\setcounter{section}{0}
\renewcommand{\theequation}{\Alph{section}.\arabic{equation}}

\section{Modified Bessel functions and elliptic integrals}
\setcounter{equation}{0}
\subsection{Modified Bessel functions}
\label{Bessel}

Here we list some useful properties (see e.g. \cite{Magn66}) of the modified
Bessel functions 
\bel{A-0}
I_n(t) = \frac{1}{2\pi}\int_{-\pi}^\pi d\phi\, \mbox{e}^{i\phi n+\cos{\phi}t}
\ee
with integer index $n$.\\
(i) Asymptotic behavior ($t\to\infty$, $u=n^2/t$ finite):
\bel{A-1}
\mbox{e}^{-t}I_n(t) \sim \frac{1}{\sqrt{2\pi t}} \mbox{e}^{-n^2/(2t)}
\ee
(ii) Recursion relations:
\bea
\label{A-2}
I_n(t) & = & I_{-n}(t)\\
\label{A-3}
2n\,I_n(t) & = & t\,(\,I_{n-1}(t) - I_{n+1}(t)\,)\\
\label{A-4}
2\,\frac{d}{dt}\,I_n(t) & = & I_{n-1}(t) + I_{n+1}(t)
\eea
(iii) Summation formulae:
\bea
\label{A-5}
\sum_{n=-\infty}^\infty \mbox{e}^{-t}I_n(t) & = & 1\\
\label{A-6}
\sum_{n=-\infty}^\infty I_n(t)I_{n+m}(t) & = & I_m(2t)
\eea
(iv)
Integrals:\\
For $m,n$ integers, one has
\begin{eqnarray}
\label{A-7}
I_m(t)I_n(t)&=&\frac{2}{\pi}\int_{0}^{\pi/2}d\theta
\cos((n\mp m)\theta)I_{n\pm m}(2t\cos\theta)\,.
\nonumber\\
\end{eqnarray}

One defines \cite{Abr70} the functions $f_{q,s}(t)$
where $q$ is a positive integer and $s$ is an integer as
\begin{eqnarray}
\label{A-8}
f_{q,s}(t)&\equiv&
\sum_{p=0}^{\infty}
\,\left(
\begin{array}{c} p+q-1\\p \end{array}\right)\,
2^q\,(-1)^p\,I_{q+s+2p}(t)\,.
\end{eqnarray}

One can also show, using the integral representation
of the modified Bessel functions, that for $s\geq 0$, one has
\begin{equation}
\label{A-9}
f_{q,s}(t)=\frac{1}{(q-1)
!}\,\int_{0}^{t}du\,I_{s}(u)\,(t-u)^{q-1}\,.
\end{equation}

The following useful equalities also hold
\begin{eqnarray}
\label{A-10}
f_{q,s}(t)&=&\frac{1}{2}\,(\,f_{q+1,s+1}(t)+f_{q+1,s-1}(t)\,)\,,
\\
\label{A-11}
f_{q+1,1}(t)&=&f_{q,0}(t)-\frac{t^q}{q!}\,,
\\
\label{A-12}
f_{q+1,-1}(t)&=&f_{q,0}(t)+\frac{t^q}{q!}\,,
\end{eqnarray}
where the first equality follows from the integral representation
of $f_{q,s}(t)$, the second follows from integration by parts of
(\ref{A-9}) and the third follows from the two above.

One can also show \cite{Prud86}, that the following identity holds
\begin{eqnarray}
\label{A-13}
\sum_{p=0}^{\infty}I_{p+m}(x)y^p&=&x\int_{0}^{1}dv\,(1-v)^m\,
e^{-\frac{1}{2}xyv^2+xyv}\nonumber\\
& &\mbox{}\times I_{m-1}(x(1-v))\,,
\end{eqnarray}
where $m$ is a positive integer.

\subsection{Elliptic integrals}

The following relations \cite{Byrd71} are used in the calculation of
Laplace transforms:
\begin{eqnarray}
\label{A-14}
\int_{0}^{\pi/2}d\theta\,
\frac{\sin^2\theta}{(1-\alpha^2\sin^2\theta)\sqrt{1-k^2\sin^2
\theta}}=\nonumber\\
\frac{\pi(1-\Lambda_0(\phi,k))}{2\sqrt{\alpha^2(1-\alpha^2)
(\alpha^2-k^2)}}\,,
\end{eqnarray}
where $k<\alpha$, $\sin\phi=\sqrt{(1-\alpha^2)/(1-k^2)}$ and
$\Lambda_0(\phi,k)$ is given in terms of elliptic functions by
\begin{eqnarray}
\label{A-15}
\Lambda_0(\phi,k)&=&\frac{2}{\pi}(\,E(k)F(\phi,k')+K(k)E(\phi,k')
\nonumber\\
& &\mbox{}-K(k)F(\phi,k')\,)\,,
\end{eqnarray}
where $K(k)$ and $E(k)$ are the complete elliptic integrals and
$F(\phi,k')$, $E(\phi,k')$ are the elliptic integrals of first and
second kind, with  $k'=\sqrt{1-k^2}$.

\section{Derivation of the exact expression for the two-point correlation
function}
\setcounter{equation}{0}
\label{Bethe}
In order to derive the exact expression for the two-point correlation
function from equation (\ref{3-6}), one needs, as stated above, to
perform the double sum $\sum_{k_0=1}^{\infty}
\sum_{l_0=k_0+1}^{\infty}$ in (\ref{3-6}) and then to determine the 
contour of integration of the double integral in this equation 
which will yield the correct initial condition, namely 
$C_{k,l}(0)=0$. In order to perform the first step, the key
point is to realize that one can write the factors $S(p_1,p_2)-1$,
$S(-p_1,p_2)-1$ and $S(-p_1,p_2)\,S(p_1,p_2)-1$ which appear
in (\ref{3-6}) in a way such that the sums over $l_0$ coming
from each of these terms can be
written as the difference of two sums starting at
neighbouring arguments, e.g. $l_0=k_0+1$ and $l_0=k_0+2$,
and can thus be easily performed using the telescopic property of
sums. Furthermore, after some tedious but straightforward
algebraic manipulations one can show that the sums over $k_0$ can
also be performed in the same way, i.e. using the telescopic property.
When performing this second sum, 
one also generates one extra term which cancels
exactly the first term of (\ref{3-6}). One obtains, after interchanging
$p_1$ and $p_2$, the following result
\begin{eqnarray}
\label{B-1}
C_{k,l}(t)&=&-2(\Delta\rho)^2\int^{'}\,\int^{'}
\,\frac{dp_1}{2\pi}\,\frac{dp_2}{2\pi}\,
e^{-(\epsilon_{p_1}+\epsilon_{p_2})t}\\
& &\mbox{}\times\left(\,
\frac{1+e^{-ip_1}}{1-2e^{-ip_1}+e^{-ip_1-ip_2}}\,
e^{-ip_2k-ip_1l}\right.\nonumber\\
& &\mbox{}\left.-\frac{1+e^{-ip_1}}{1-2e^{-ip_1}+e^{-ip_1+ip_2}}\,
e^{-ip_2k-ip_1l}\,\right)\,.\nonumber
\end{eqnarray}

Now we need to determine the appropriate contour 
of integration in (\ref{B-1}). But in fact, one 
does not need to determine it explicitly. Assuming that the
contour includes the origin, one can 
use the identity $1/x=\int_{0}^{\infty}\, d\alpha\,
e^{-\alpha x}$ to represent each of the denominators of
(\ref{B-1}) as an integral over $\alpha$
and then formally expand the resulting exponentials under
the integration sign in powers of $e^{-ip_1}$ and $e^{-ip_1-ip_2}$
for the first denominator, $e^{-ip_1}$ and $e^{-ip_1+ip_2}$
for the second denominator. If one then performs the integrals over
$p_1$, $p_2$ and $\alpha$, one obtains the following 
result for $C_{k,l}(t)$
\begin{eqnarray}
\label{B-2}
C_{k,l}(t)&=&-(\Delta\rho)^2\,e^{-2t}\,\sum_{p,q=0}^{\infty}
\,\left(
\begin{array}{c} p+q\\p \end{array}\right)\,
2^{q+1}\,(-1)^p\\
& &\mbox{}\times\left(\,
I_{p+k}(t)\,I_{p+q+l}(t)+I_{p+k}(t)\,I_{p+q+1+l}(t)\right.\nonumber\\
& &\mbox{}-\left.
I_{p-k}(t)\,I_{p+q+l}(t)-I_{p-k}(t)\,I_{p+q+1+l}(t)\,\right)\,.
\nonumber
\end{eqnarray}
Notice that the existence of the expansions depends on the
convergence of the resulting series, which implicitly
fixes the contour.
It can be easily checked from the properties
of the modified Bessel functions that this expression
does indeed obey the initial condition
$C_{k,l}(0)=0$ (notice that $k<l$).

This expression is still rather cumbersome to use. If we
apply (\ref{A-7}) to products of two modified Bessel functions,
we will obtain, with $m=l-k$, $n=l+k$, $r=q+2p$
\begin{eqnarray}
\label{B-3}
C_{k,l}(t)&=&-\frac{2(\Delta\rho)^2}{\pi}
e^{-2t}\!\!\int_{0}^{\pi/2}d\theta
\sum_{p,q=0}^{\infty}
\left(
\begin{array}{c} p+q\\p \end{array}\right)2^{q+1}(-1)^p
\nonumber\\
& &\mbox{}\times 
\left(\,I_{r+1+(n-1)}(\,2t\cos\theta\,)\,\cos(\,(q+m)\theta\,)\right.
\nonumber\\
& &\mbox{}+I_{r+1+n}(\,2t\cos\theta\,)\,\cos(\,(q+1+m)\theta\,)
\nonumber\\
& &\mbox{}-I_{r+1+(m-1)}(\,2t\cos\theta\,)\,\cos(\,(q+n)\theta\,)
\nonumber\\
& &\mbox{}\left.-I_{r+1+m}(\,2t\cos\theta\,)\,\cos(\,(q+1+n)\theta\,)\,\right)
\,.
\end{eqnarray}

One can now write (\ref{B-3}) in terms of the $f_{q,s}(t)$
functions which were defined above. One has
\begin{eqnarray}
\label{B-4}
C_{k,l}(t)&=&-\frac{2(\Delta\rho)^2}{\pi}
\,e^{-2t}\int_{0}^{\pi/2}d\theta\\
& &\sum_{q=0}^{\infty}
\left[\,f_{q+1,l+k-1}(2t\cos\theta)\,\cos(\,(q+l-k)\theta\,)
\right.\nonumber\\
& &\mbox{}+
f_{q+1,l+k}(2t\cos\theta)\,\cos(\,(q+l-k+1)\theta\,)
\nonumber\\
& &\mbox{}-
f_{q+1,l-k-1}(2t\cos\theta)\,\cos(\,(q+l+k)\theta\,)
\nonumber\\
& &\mbox{}-\left.
f_{q+1,l-k}(2t\cos\theta)\,\cos(\,(q+l+k+1)\theta\,)\,
\right]\,.\nonumber
\end{eqnarray}

Using the integral representation (\ref{A-9}) and summing
over $q$ one obtains, after the substitution $u=2(t-v)\cos\theta$
in the integral over $u$, the solution (\ref{3-7}) where
the functions $F_{m,n}(x)$ are given by (\ref{3-8}).

Since the steps which led from (\ref{B-1}) to (\ref{B-2})
are only formal, one should check explicitly that (\ref{3-7})
is indeed a solution of the equations of motion for
the joint probabilities $\exval{n_{k}(t)n_{l}(t)}$.
This is trivial for $l\neq k+1$. For $l=k+1$ one
can show that the unphysical amplitudes 
$C_{k,k}(t)$, $C_{k+1,k+1}(t)$ which are
generated by the time derivative
of $\exval{n_{k}(t)n_{k+1}(t)}$, obey the following
identity
\begin{eqnarray}
\label{B-5}
C_{k,k}(t)+C_{k+1,k+1}(t)-2C_{k,k+1}(t)=\\
-(\Delta\rho)^2
\,e^{-2t}\,(I_{k}(t)+I_{k+1}(t))^2\,,\nonumber
\end{eqnarray}
which cancels exactly the unphysical contribution
coming from the term $(\rho_{k}(t)-\rho_{k+1}(t))^2$
(see equations (\ref{3-1}) and (\ref{3-1c}))
which also appears in the equation for
$\exval{n_{k}(t)n_{k+1}(t)}$, thus showing
that $\exval{n_{k}(t)n_{k+1}(t)}$ obeys the
correct equation of motion.
This identity can be proved by considering the
expression for the lhs of (\ref{B-5}) as given
in terms of the integral representation (\ref{3-7}).
After some cancelations between the terms, 
one uses the identity
\begin{eqnarray}
\label{B-6}
\frac{d}{dv}(\,e^{v(1+\cos(2\theta))}\,
\cos(v\sin(2\theta)+s\theta)\,)=\nonumber\\
2\,e^{v(1+\cos(2\theta))}\,
\cos(v\sin(2\theta)+(s+1)\theta)\,\cos\theta
\end{eqnarray}
to integrate the resulting expression by parts. Applying
the identity (\ref{A-7}) to the remaining
integrals, in order to transform them into products of modified
Bessel functions yields the desired result. An alternative
route to derive (\ref{B-5}) is to use the representation
(\ref{B-4}) and the equalities (\ref{A-10}) to (\ref{A-12}).

\section{Asymptotics of $\sigma^2$}
\setcounter{equation}{0}
\label{Laplace}

\subsection{Sum over $g_k^2(t)$}

This expression arises in the summation over $\exval{n_k(t)}^2$ which
forms part of the `dynamical compressibility' $\exval{N^2(t)}-
\exval{N(t)}^2$ entering the expression for $\sigma^2(t)$.
It is convenient to split this sum into three different parts. One has
\bea
\sum_{k=1}^\infty g_k^2(t) & = & -  \mbox{e}^{-2t} \sum_{k=1}^\infty I_k^2(t)
- 2 \mbox{e}^{-2t} \left(\sum_{k=1}^\infty I_k(t)\right)^2 \nonumber \\
& & + 4 \mbox{e}^{-2t}\sum_{k=1}^\infty\sum_{p=k}^\infty\sum_{q=k}^\infty
I_p(t)I_q(t).
\eea
The first part can be evaluated using (\ref{A-2}), (\ref{A-6}), the second 
part using (\ref{A-5}). To evaluate the third part we rewrite the summation
\be
\sum_{k=1}^\infty\sum_{p=k}^\infty\sum_{q=k}^\infty I_p(t)I_q(t) =
\sum_{k=1}^\infty \left(k\,I^2_k(t)+2\sum_{p=0}^{k-1}p\,
I_p(t)I_k(t)\right)
\ee
and apply (\ref{A-2}), (\ref{A-3}), (\ref{A-5}), (\ref{A-6}). One obtains
\bea
\sum_{k=1}^\infty\sum_{p=k}^\infty\sum_{q=k}^\infty I_p(t)I_q(t) &=&
\frac{t}{2}\left[\mbox{e}^{t}(I_0(t)+I_1(t))\right.\nonumber\\
&&\mbox{}\left. -(I_0(2t)+I_1(2t))\right].
\eea
Putting everything together yields 
\bea
\sum_{k=1}^\infty g_k^2(t) & = & 2t\left[\mbox{e}^{-t}(I_0(t)+I_1(t))-
\mbox{e}^{-2t}(I_0(2t)+I_1(2t))\right] \nonumber \\
 & & -(1-2\mbox{e}^{-t}I_0(t)+\mbox{e}^{-2t}I_0(2t))/2.
\eea
The second part of this expression is subleading in time and can be
ignored in the study of the asymptotic behavior.

\subsection{Laplace transform of $K(t)$}
In order to perform the sum on the right hand side of 
(\ref{4-4a}), we consider the Laplace transform of the 
correlation function $C_{k,l}(t)$ as given by (\ref{3-7}).
Since the expression (\ref{3-7}) involves a convolution
of two functions $\int_{0}^{t}dv\,f(t-v)g(v)$, 
such transformation simplifies considerably the calculations,
because the Laplace transform of such a convolution is the
product of the Laplace transforms of the two functions.
Writing the Laplace transform as $\tilde{C}_{k,l}(s)$,
one has
\begin{eqnarray}
\label{CL-1}
\tilde{C}_{k,l}(s)&=&-\frac{4(\Delta\rho)^2}{\pi}\,\int_{0}^{\pi/2}
d\theta\,\frac{\cos\theta}{\sqrt{(s+2)^2-4\cos^2\theta}}\nonumber\\
& &\mbox{}\times\frac{1}{((s+1)^2-2(s+1)\cos(2\theta)+1)}\nonumber\\
& &\mbox{}\times
\left[ {\cal F}_{k+l-1,l-k}(s) +
{\cal F}_{k+l,l-k+1}(s) \right.\nonumber \\
& &\left.\mbox{}-{\cal F}_{l-k-1,k+l}(s)-{\cal F}_{l-k,k+l+1}(s)\right]\,,
\end{eqnarray}
where the functions ${\cal F}_{m,n}(s)$ are defined by
\begin{eqnarray}
\label{CL-2}
{\cal F}_{m,n}(s)&=&\left(\frac{s+2-\sqrt{(s+2)^2-4\cos^2
\theta}}{2\cos\theta}\right)^m\nonumber\\
& &\mbox{}\times [(s+1)\cos(n\theta)-\cos((n-2)\theta)]\,.
\end{eqnarray}

An additional advantage
of (\ref{CL-1}) with respect to (\ref{3-7}) is that
the sums over $k$ and $l$ now reduce to summing two
geometric series, due to the form of ${\cal F}_{m,n}(s)$.
Performing such sums, one obtains after computing some
standard integrals using the residue theorem, the following expression
for the Laplace transform $\tilde{K}(s)$ of $K(t)$
\begin{eqnarray}
\tilde{K}(s)&=&-\frac{2}{s^{3/2}}\left(\sqrt{s+2}
-\frac{\sqrt{s+4}}{4}\right)
+\frac{3}{2s}\\
& &\mbox{}+\frac{4(3s+4)}{\pi(s+2)^3}\nonumber\\
& &\mbox{}\times
\int_{0}^{\pi/2}d\theta
\frac{\sin^2\theta}{(1-\alpha^2(s)\sin^2\theta)\sqrt{1-k^2(s)\sin^2
\theta}}\nonumber\\
& &\mbox{}-\frac{8}{\pi(s+2)^2}\nonumber\\
& &\mbox{}\times
\int_{0}^{\pi/2}d\theta
\frac{\sin^2\theta}{(1-\alpha'^2(s)\sin^2\theta)\sqrt{1-k^2(s)\sin^2
\theta}}\,,\nonumber
\end{eqnarray}
where $\alpha^2(s)=\frac{4(s+1)}{(s+2)^2}$,
$\alpha'^2(s)=\frac{2}{s+2}$ and $k^2(s)=\frac{4}{(s+2)^2}$.
Since the last two terms are of the form (\ref{A-14}),
one can now expand the elliptic integrals 
at small $s$ \cite{Byrd71}. The most singular terms of
this expansion, i.e. the terms which diverge
as $1/s^{3/2}$  at small $s$ diverge like $\sqrt{t}$ at large $t$
in the time-domain, i.e. after inverting the Laplace transformation.
Collecting all these leading order terms in the expansion of $\tilde{K}(s)$
yields the result given in equation (\ref{4-5a}).

\end{document}